\documentclass[12pt]{article}
\usepackage{epsfig}
\def\beq{\begin{equation}}
\def\eeq{\end{equation}}
\def\ben{\begin{eqnarray}}
\def\een{\end{eqnarray}}
\def\bea{\begin{array}}
\def\eea{\end{array}}

\begin{document}

\baselineskip=20pt


\begin{flushright}
Prairie View A \& M, HEP-3-96\\
July 1996 \\
Los Alamos National Lab bulletin board: hep-ph/9612271

\end{flushright}

\vskip.15in

\begin{center}
{\Large  FCNC, CP violation, and
 impure Majorana neutrinos
}

\vskip .3in

{\bf Dan-Di Wu}\footnote{ $~$E-mail:  wu@hp75.pvamu.edu or 
danwu@physics.rice.edu }
\vskip.6in

{\sl HEP, Prairie View A\&M University, Prairie View, 
TX 77446-0355, USA}
\end{center}
\vskip.36in

{\sl The tree level diagonalization of a neutrino mass matrix with both 
Majorana and Dirac masses is discussed in a general context. Flavor changing
neutral currents in such models are inevitable. Rephasing invariant 
quantities characterizing CP violation in FCNC Fermion-Higgs interactions 
are identified. At the one loop level, the mass eigenstates become an impure 
Majorana type. The possibility of a significant change in the mass spectrum 
for the left-handed neutrinos is explored, with an example of two species 
of neutrinos. Neutrino oscillations with impure Majorana neutrinos are also 
discussed.
 }

\vskip 2.5in

\clearpage

\vskip.1in
Data from more and more neutrino oscillation experiments 
hint at non-zero neutrino masses[1]. 
Furthermore, the available data seem to indicate that 
the mass squared
differences of the 
neutrinos, if they exist,  form a hierarchical spectrum. 
This is to be compared with the fact that  the masses of each type 
of charged
fermions have a hierarchical spectrum. 
If neutrino oscillations exist, this will be a very strong 
motivation for introducing neutrino masses in the theory.
Another motivation for massive neutrinos is that they
could be a candidate
 for the dark matter in the universe.   
\vskip.1in
Neutrinos are very different from other fermions, 
because it is electrically neutral.  
Since neutrinos are neutral, it may have two different types 
of masses, 
the Dirac masses, which is similar to the masses of all 
charged fermions; 
and the Majorana masses, which require the violation of the 
fermion number conservation. Since fermion number conservation is 
not regarded 
a fundamental principle in most gauge interaction models,
neutrinos are very likely to have Majorana masses, if they 
are massive. 
\vskip.1in
In the minimal standard model (SM) 
 neutrinos are massless. There is no way to write a mass term 
 for the neutrinos
in SM without breaking the gauge symmetry of this model.  
 However, one can
 give neutrinos masses with a minimal extension of the SM.
One can, for example, add a Higgs triplet $\Delta_L$ to SM 
and obtain
a new  term
\beq
\kappa^L_{ij} \psi^T_{_{Li}}C \Delta_L \psi_{_{Lj}}  +h.c.
\eeq
where $\psi_{_L}=\left(\bea{c}
\nu_{_L}\\
e_{_L}\eea\right)$ is the lepton doublet. Note, $\nu_L$ here as an 
eigenstate of interactions is a Weyl spinor. $\kappa^L_{ij}$ 
is a $3\times 3 $
matrix of the Majorana
 coupling constants. This matrix must be symmetric, because 
$\psi^T_{_{Li}}C  \psi_{_{Lj}}= \psi^T_{_{Lj}}C  \psi_{_{Li}}$. The anti-symmetric 
part of the 
coupling matrix, if it exists, does not contribute 
(The anti-symmetric part is cancelled out by
itself.). If the vacuum expectation value of 
$\Delta_L$ is non-zero, then neutrinos will be massive, 
with a symmetric mass matrix $M_L$, and the lepton number 
conservation
 will be spontaneously broken.
One needs both small  coupling constants and small VEV of 
$\Delta_L$ to 
accommodate for the observed tiny neutrino masses in such a model. 
\vskip.1in
One can also introduce, instead of $\Delta_L$,  
right-handed neutrinos $\nu_{_{Ri}}$. 
One then has
\beq
g_{ij}\nu_{_{Ri}}^{cT}C\phi^T\psi_{_{Lj}}
+\kappa^R_{ij}\mu\nu_{_{Ri}}^{cT}C\nu_{_{Rj}}^c +h.c.
\eeq
where $\kappa^R_{ij}$ is a symmetric Majorana coupling matrix,
 and $\mu$ is a mass scale.
Here again we meet with  Majorana mass terms, the mass 
terms for the 
right-handed neutrinos.
  These mass terms do
not vanish, unless fermion number conservation is imposed 
on the model. The Dirac-type Yukawa
coupling constants $g_{ij}$ are an arbitrary matrix.
A possible criticism to this model says  that since the fields 
$\nu_{_{Ri}}$
do not have any gauge interactions of the SM, 
 it does not look appealing to have 
them; unless there are some other interactions beyond the SM 
interactions
which involve $\nu_{_{Ri}}$. After $\phi$ develops VEV, we obtain a 
$6\times 6$ symmetric mass matrix, if there are three 
generations of fermions.
The mass matrix of mass terms (2) is, in the form of 
$2\times 2$ block matrix 
\ben
\left(\bea{cc}
0&M\\
M^T&M_R\eea\right)
\een
where $M_{ij}=\frac{1}{2}g_{ij}v$, and $M_{Rij}=\kappa^R_{ij}\mu$.
 When $\mu=0$ this 
is the model with only Dirac masses.
\vskip.1in

Combining mass terms
 (1) and (2), a general mass matrix of the neutrinos has the form
\ben
\left(\bea{cc}
M_L&M\\
M^T&M_R\eea\right)
\een
which is symmetric[2], with $M_{Lij}=\kappa^L_{ij}<\Delta_L>$.
\vskip.1in

 Models like the left-right symmetric models,  
 the $SO(10)$ grand unification models[3], and their 
  supersymmetric extensions,   necessarily require massive 
  neutrinos. 
A characteristic property of these models is that they all need 
neutrinos to have Majorana masses, if neutrino masses of the 
order of 
the masses of the charged fermions are to be avoided. Indeed, 
in these models,
neutrinos obtain Dirac masses which are 
``naturally" compatible with the masses of 
charged fermions.  In order to 
produce the acceptable tiny masses for left-handed neutrinos, 
out of these much 
too large Dirac neutrino masses, 
a see-saw mechanism  must be
applied[2]. Consequently, one needs at least an original large 
right-handed 
Majorana neutrino  mass matrix. The effective neutrino mass terms 
in 
these models  can be represented by Eq (2), or by the 
combination of Eqs (1) and
(2). The latter case with a non-zero VEV for $\Delta_L$
 is not popular. In either case, it is not difficult to make the 
model  satisfy the condition $\kappa^L_{ij}=\kappa^R_{ij}$ 
by suitably 
applying the L - R symmetry of the models.
\vskip.1in

This note will be divided into two parts. The first part is limited
 to the tree-level discussions. Flavor changing neutral 
 currents (FCNC) 
interactions mediated by Higgs bosons
 and neutrino related CP violation are discussed in a general 
 context. 
First loop effects will be discussed in the second part. These 
effects include 
impure Majorana states as mass eigenstates and the asymmetry of 
decay products 
due to CP violation. The potential of
a significant change in the mass
splitting between  Majorana neutrinos is explored. 
\vskip.1in
First let us discuss 
the general diagonalization problem in order to see the FCNC 
neutrino 
interactions on a precise basis. 
A symmetric  matrix ${\bf\large S}$, like that in Eq (3) or (4),
 can be diagonalized by one unitary matrix in a symmetric way[4],
\beq
U^T {\bf\large S}U=D
\eeq
where $D$ is a diagonalized matrix with all elements real and 
no less than zero,
 and
$U=U_1^T\times e^{i\phi}$ with  $U_1$, a unitary matrix which 
satisfies
$$U_1{\bf\large SS}^\dagger U_1^\dagger={\bf\large D}^2;$$
and $\phi $ is a real diagonal matrix such as to make $D$ real and 
definitely 
non-negative.
\vskip.1in

If the see-saw mechanism is at work, block diagonalizing the 
complete neutrino 
mass matrix of the type of Eq (3) with $M_R >> M$,
one   ends up with a tiny  Majorana
mass matrix for the left-handed neutrinos, as expressed in 
the following 
formula[2]

\beq
M^{NL} \dot{=} - MM_{R}^{-1}M^{T}
\eeq
where $M_{R}$ is the large $3\times 3$ right-handed 
Majorana mass matrix,  
and $M$ is the  $3\times 3$ Dirac mass matrix of the neutrinos.
  Here we assume the original 
left-handed Majorana 
mass matrix is either zero or negligible ($M_L=0$). Otherwise 
there will be 
an additional $3\times 3$ 
original left-handed Majorana mass matrix at the right-hand 
side of the equation. Note that Eq (6) is symmetric, which is 
consistent with
its Majorana property. This formula is widely used in the 
literature, 
but sometimes wrongly recorded. 

\vskip.1in
Second, one notices that
 the flavor changing neutral currents (FCNC) in the neutrino sector 
is inevitable and copious, when Dirac and Majorana 
masses are both present in the 
models,  particularly in any see-saw models of neutrino masses.
 Indeed, expressing the unitary matrix which diagonalizes the 
 full neutrino mass matrix 
in a $2\times 2 $ block form
\ben
U=\left(\bea{cc}
U_L^N &U_{12}\\
U_{21}&U^N_R   \eea\right),
\een
one finds that the following matrices are diagonalized  
very  accurately
after the full mass diagonalization 
\beq
M_R^D = U^{NT}_RM_RU^{N}_R, 
\eeq
and\footnote{Or $U_L^{NT}M^{NL}U_L^N +U_L^{NT}M^{L}U_L^N$ if 
$M_L$ in (4) is
tiny but still significant.}
\beq
M_L^D = U_L^{NT}M^{NL}U_L^N 
\eeq
where $M^{NL} $ is defined in Eq (6).
However the new Majorana coupling constants between the 
left-handed 
neutrinos and 
the left-handed triplet $\Delta_L$
\beq
\kappa^N=U_L^{NT}\kappa_LU_L^N=U_L^{NT}U^{N*}_R\frac{M_R^D}{V_R}
U_R^{N\dagger}U_L^N
\eeq
is not diagonalized, where left-right symmetry is assumed in 
the second step.
 Neither is
 the new Yukawa coupling constants among the left-handed neutrinos, 
right-handed neutrinos and 
 the doublet Higgs $\phi^0$
\beq
g^N=U_L^{N\dagger}gU^N_R.
\eeq
 Consequently, flavor changing neutral 
currents (FCNC) mediated by the neutral component of the triplet
 $\Delta^0$ and by the neutral 
component of the doublet $\phi^0$ exist in general. 
\vskip.1in

One may wonder how these FCNC interactions have
affected the abundance of different 
species of relic neutrinos from the big-bang universe. 
A quick examination tells that these interactions are too weak to
 have such an effect. Actually, 
the average  rate of the relevant process (e.g. $\nu_{_{L3}} 
+ \nu_{_{L1}}
\rightarrow \Delta^0_L\rightarrow 2\nu_{_{L1}}$) $\langle 
\sigma v n\rangle_{T_d}$
at the temperature $T_d\sim $ 10 eV, the assumed heaviest mass of 
left-handed neutrinos, is much smaller than the then 
Hubble  constant $H(T_d)$ for all reasonable masses of 
$\Delta^0_L$.
\vskip.1in  
As 
CP violation involved in massive neutrino
models is concerned, one notices that CP violation
 comes from several sources. There are CP violations mediated by 
$W_{L}$ and $W_{R}$, which are characterized by 
their corresponding CKM matrices.
Of special interest is 
  CP violation  rooted in the non-diagonal 
neutrino interactions of (10) and (11). 
CP violation in Higgs coupling constants is defined by the imaginary
parts of the
rephasing invariant quartets[5], e.g.
\beq
\Delta^\kappa_{i\alpha}=\varepsilon_{ijk}
\varepsilon_{\alpha\beta\gamma}
\kappa^N_{j\beta}\kappa^N_{k\gamma}
\kappa^{N*}_{j\gamma}\kappa^{N*}_{k\beta}\hskip.2in ({\rm no\ summation})
\eeq
It is found that it is impossible to make a non-trivial 
quartet by interfering
two tree diagrams for the Majorana couplings in Eq (10).
 Therefore,
tree level CP violation with FCNC Majorana couplings does not exist.
Such CP violation exists for  the Yukawa couplings in Eq (11), 
which is 
quite similar to the charged current gauge couplings, 
except for the lack of universality for the Yukawa couplings.
Note, since $\kappa^N$ (or $g^N$)  here is not  unitary, therefore 
Im$\Delta^\kappa_{i\alpha}$ (Im$\Delta^N_{i\alpha}$) 
for different processes are different. These quantities are small if 
there is a hierarchy in  the matrix elements. Since all the 
phases in the 
coupling matrices are subject to change by rephasing the 
neutrino fields,
there are only four independent useful
phases in $g^N$ and three in $\kappa^N$, if 
these matrices are of dimension $3\times 3$. Consequently, when one 
$\Delta_{i\alpha}$ is
purely imaginary (so-called having a maximal CP violation), 
the others may not. 
Such a concept of maximal CP violation is widely used in 
the estimation of baryon 
excess due to CP violation[6].
The number of pure imaginary $\Delta_{i\alpha}$ is limited to three
for $\kappa^N$ and four for $g^N$ respectively.
One can also find a matrix with all its $\Delta_{i\alpha}$ having 
significant
phases,
for instance, a matrix with the following phase distribution: 
$$\left(\bea{ccc}
0        &\sqrt{ i}&\sqrt{ i}\\
\sqrt{ i}&0        &\sqrt{ i}\\
\sqrt{ i}&\sqrt{ i}&0          \eea\right), 
$$
where a zero element means the matrix element is real.
 This is to be compared with maximal CP violation 
in the CKM matrix, where none of the quartets can be made 
purely imaginary
because of the unitarity constraint.
\vskip.1in
Other interesting new CP violation sources are in the charged 
Majorana 
couplings. For example, the $\Delta_L^-$ coupling of Eq (1)
\beq
\l^T_{_{Li}}C\ \ _{_{_i}}\left(U^{lT}_L\kappa^L U^{N}_{L}
\right)_j\Delta_L^+ \nu_{_{Lj}} 
+ {\rm transposed + h.c.}
\eeq
There will be a similar term for the right-handed  neutrinos, 
if the model is L-R symmetric, and there is only one pair of 
$\Delta_L$ -
$\Delta_R$,
\beq
\l^T_{Ri}C\ \ _{_{_i}}\left(U^{lT}_{R}U_R^{N*}\frac{M^D_R}
{V_R}\right)_j\Delta_R^+ \nu_{_{Rj}} 
+ {\rm transposed + h.c.}
\eeq
Of course each of these coupling constants
 have their own corresponding quartets to be discussed. 
\vskip.1in
The above discussions are only valid at the tree level. When loop 
effects are introduced, there will be some more interesting 
physics, in particular,
the physics somehow resembles  that in the $K^0 -\bar{ K^0}$ system.
This piece of physics, especially that of  
right-handed neutrinos, has recently aroused some enthusiasm due to
a brilliant paper by Flanz, Paschos, Sarkar, and Weiss (FPSW)[7].
Essentially,  the Weyl fields $\nu_{_R}$ and $\nu_{_R}^c$, which are 
the eigenstates
of interactions, are mutually CPT conjugated.
 This
system would have been an exact copy of the $K^0-\bar{K^0}$ 
system with
 asymmetric decay products (e.g. the $l^+/l^-$ ratio is not 1), 
if  mixing mass terms between $\nu_{_R}$ and $\nu_{_R}^c$ were not 
forbidden by Lorentz invariance. 
FPSW  found a two species system, which may fulfill that 
kind of mixing. 
 A further study of their system will be given here.
 An extension of their discussion to the left-handed
neutrinos will  also be attempted.
\vskip.1in
Consider first  two neutrino states of the same chirality, the 
Hamiltonian at the tree level after mass diagonalization 
is $(\nu_{_R}\equiv \nu$,
and assuming CPT)             
\ben
(\nu_1^c, \nu_2^c, \nu_1, \nu_2)\hat{\it H}^0\left(\bea{c}
\nu_1\\
\nu_2\\
\nu_1^c\\
\nu_2^c
\eea\right),
\een
where 
\ben
\hat{\it H}^{(0)}=\left(\bea{cccc}
0    &0    &M_{11}&0   \\
0    &0    &0  &M_{22} \\
M^*_{11}&0    &0  &0   \\
0    &M^*_{22}&0  &0
\eea\right).
\een
Note here only neutrinos with a specific chirality (e.g. $\nu_{_R}$
and $\nu_{_R}^c$,
instead of $\nu_{_L}^c$ ) are considered. 
This form is convenient for separating 
the absorptive part from the dispersive part of the Hamiltonian,
which will become clear in a moment.
At the tree level, the Weyl and the Majorana states are 
equivalent, 
so far as the mass (and the decay life time) 
eigenstates  are concerned.
This is not true when loop effects are considered. 
The loop corrections (see Fig. 1, which is copied from Ref[7].)
to the zeros in $\hat{\it H}^{(0)}$ are convergent[8],
if the theory is renormalizable and these 
corrections do not have counter terms. Including loop effects, the
total Hamiltonian is expressed as
\ben
\bea{ccc}
\hat{\it H}&=&\hat{\it H}^{(0)}+ \hat{\it H}^{(corr)} = 
\left(\bea{cccc}
0                  &0                  &\alpha_{11}&\alpha_{12}\\
0                  &0                  &\alpha_{12}&\alpha_{22}\\
\tilde{\alpha}_{11}&\tilde{\alpha}_{12}&0          &0          \\
\tilde{\alpha}_{12}&\tilde{\alpha}_{22}&0          &0
\eea\right)\hskip.5in\\
&&\\
           &=&
\left(\bea{cccc}
0                                 &0                               
 &
     M_{11}-\frac{i}{2}\Gamma_{11}&M_{12}-\frac{i}{2}\Gamma_{12}\\
0                                 &0                               
 &
     M_{12}-\frac{i}{2}\Gamma_{12}&M_{22}-\frac{i}{2}\Gamma_{22}\\
M^*_{11}-\frac{i}{2}\Gamma^*_{11} &M^*_{12}-\frac{i}{2}
\Gamma^*_{12}&0&0\\
M^*_{12}-\frac{i}{2}\Gamma^*_{12} &M^*_{22}-\frac{i}{2}
\Gamma^*_{22}&0&0
\eea\right).
\eea
\een
There are no odd terms of the eigenvalue $\lambda$ in the 
eigenequation  for $\hat{\it H}^{(0)}
+ \hat{\it H}^{(corr)}$. One therefore has 
\beq
\lambda_2=-\lambda_1, \hskip.2in \lambda_4=-\lambda_3.\hskip.3in 
({\rm Im}\lambda_1\le 0,\ \
{\rm Im}\lambda_3\le 0)
\eeq
The eigenvectors turn out to be almost Majorana
states, or impure Majorana states as they are called. These eigenstates
$M_\beta=(x_\beta,y_\beta,z_\beta,w_\beta)$, with 
$\beta = 1,2,3,4$,  are expressed as, up to normalization constants 
\ben
\bea{cccc}
x_\beta&=& &(\alpha_{11}\alpha_{22}-\alpha_{12}^2)
             \tilde{\alpha}_{12}+\lambda_\beta^2\alpha_{12},\\
y_\beta&=&-&(\alpha_{11}\alpha_{22}-\alpha_{12}^2)
             \tilde{\alpha}_{11}+\lambda_\beta^2\alpha_{22},\\
z_\beta&=& &\lambda_\beta(\alpha_{22}\tilde{\alpha}_{12}+
             \alpha_{12}\tilde{\alpha}_{11}), \hskip.3in    \\
w_\beta&=&-&\lambda_\beta(\alpha_{11}\tilde{\alpha}_{11}+
             \alpha_{12}\tilde{\alpha}_{12})+\lambda^3_\beta.
\eea
\een
A phase redefinition of the eigenstates with
$\beta=2, 4$, which are almost CP odd, will change the signs
of their eigenvalues, so to let them  have positive widths.
In other words, $M_1$ and $M_2$ are two orthogonal mass eigenstates
with exactly the same mass and width.
In the following we will take a phase convention
to make $\tilde{\alpha}_{11}=\alpha_{11},$$\,\,
\tilde{\alpha}_{22}=\alpha_{22}$.  

\vskip.1in
It is easy to discuss the solutions in two special cases.  

{\bf Case 1:}  $|\alpha_{22}| >> |\alpha_{11}|>>$$
|\alpha_{12}|$.  
One finds, $\lambda_1 \approx \alpha_{22}$, $\lambda_3 \approx
\alpha_{11}$, 
and to the leading orders ($\lambda_3$ must be calculated to the 
next leading orders in order to obtain the following answer.), 
the eigenvectors are: ($\delta=\alpha_{12}/\tilde{\alpha}_{22}$, 
$\tilde{\delta}
=\tilde{\alpha}_{12}/\alpha_{22}$, 
$\gamma=\alpha_{12}/\tilde{\alpha}_{12}$)  
\ben
\bea{cccccccc}
M_1&\sim& (& \delta& 1           &\tilde{\delta} &1       &),   \\
M_2&\sim& (&-\delta&-1           &\tilde{\delta} &1       &),   \\
M_3&\sim& (&1      &-\gamma\delta&1              &-\delta &),   \\
M_4&\sim& (&-1     & \gamma\delta&1              &-\delta &).
\eea
\een
Note that the mass (decay) eigenstates are of impure Majorana type. 
Assuming  charged leptons in $\nu_i$ decays while an 
equal amount of
anti-charged leptons
in $\nu_i^c$ decays are found, one then has the lepton-anti-lepton asymmetry
in the decays of these impure Majorana particles:
\ben
\bea{ccccc}
\delta_1&=&{\Gamma (M_1\rightarrow l^-+x)-\Gamma (M_1\rightarrow 
l^++\bar x)
   \over  \Gamma (M_1\rightarrow l^-+x)+\Gamma (M_1\rightarrow 
l^++\bar x)}
        &=&{|\tilde{\alpha}_{12}|^2-|\alpha_{12}|^2 \over 2|
\alpha_{22}|^2}
         =\frac{{\rm Im} M_{12}\Gamma^*_{12}}{M_{22}^2+
         \Gamma_{22}^2/4},\\
&&&&\\
\delta_3&=&{\Gamma (M_3\rightarrow l^-+x)-\Gamma_1(M_3
\rightarrow l^++\bar x)
   \over  \Gamma (M_3\rightarrow l^-+x)+\Gamma_1(M_3
\rightarrow l^++\bar x)}
	      	&=&\delta_1.
\eea
\een
The formula of $\delta_1$ can be further expressed by nontrivial 
$\Delta_{i\alpha}$ as illustrated in Ref[4, 1986]. 
\vskip.1in

{\bf Case 2:} $|\alpha_{12}|>>$$|\alpha_{22}|>>$$ |\alpha_{11}|$. \\
 In this case, the $\Gamma$ part of the Hamiltonian is negligible.
The interesting new physics is twofold: First, the masses
are enhanced from $\alpha_{11}$ and $\alpha_{22}$ to about 
$\alpha_{12}$.
Second, the splitting 
is enhanced  from $\alpha_{22}$ to $\sqrt{\alpha_{22}\alpha_{12}}$. 
Since both masses are now of the order of $\alpha_{12}$,  
this is a fascinating mechanism to obtain almost degenerate masses
(in terms of their mass ratio being close to 1) and a large mixing.
 It seems that this
mass spectrum is not favored by the present data, although the 
present data
are still to be clarified. 
The mass eigenstates are now
\ben
\left(\bea{c}
M_1\\
M_2\\
M_3\\
M_4  \eea\right)\approx\frac{1}{2}
\left(\bea{cccc}
 1& 1& 1& 1\\
 1& 1&-1&-1\\
-1& 1& 1&-1\\
-1& 1&-1& 1  \eea\right)
\left(\bea{c}
\nu_1\\
\nu_2\\
\nu^c_1\\
\nu^c_2
\eea\right),
\een
which are pure Majorana states, if the decay rates are neglected.
Looking at the $4\times 4$ full mixing, one
 may wonder whether it  necessary to work on a $4 \times 4$ 
 mixing matrix 
when discussing neutrino oscillations of two species of neutrinos.
To discuss this, let the gauge couplings be, in the $4 \times 4$ form,
\ben
\left(     \bar l_{L_1}W^+\ \ \bar   l_{L_2}W^+
     \ \ \bar l^c_{L_1}W^-\ \ \bar l^c_{L_2}W^-\right) {\large V}
\left(\bea{c}
M_{1}\\
M_{2}\\
M_{3}\\
M_{4}\eea\right)
\een
where the $4 \times 4$ mixing matrix ${\large V}$
is
\ben
{\large V}=\frac{1}{2}
\left(\bea{cccc}
 \rho_+\ &\  \rho_+\ &\ -\rho_-\ &\ -\rho_-\\
 \rho_-\ &\  \rho_-\ &\  \rho_+\ &\  \rho_+\\
 \rho_+\ &\ -\rho_+\ &\  \rho_-\ &\ -\rho_-\\
 \rho_-\ &\ -\rho_-\ &\ -\rho_+\ &\  \rho_+\\
\eea\right), \hskip.2in
(\rho_\pm=\cos\theta\pm\sin\theta)
\een
where $\theta$ is the tree level 
Cabbibo angle between the two species. Suppose at the 
production point a neutrino is produced together with $l^+_{1}$, 
then its wave function at a later time $t$  will be
\ben
\bea{ccc}
\psi(t,\nu_1)&=&\frac{1}{2}[(\rho_+M_1 + \rho_+M_2)e^{-im_1t} - 
                            (\rho_-M_3 + \rho_-M_4)e^{-im_3t}]\\
             & &					      \\
             &=&\frac{1}{2}[\rho_+(\nu_{1}+\nu_2)e^{-im_1t}+
                            \rho_-(\nu_{1}-\nu_2)e^{-im_3t}].
\eea
\een
One then has, at the detector,  
\beq
|\langle\psi(t,\nu_1)|l_{1,2}\rangle|^2=\frac{1}{2}\{1\pm
\cos2\theta
\cos[(m_1-m_3)t]\}.
\eeq
A special situation is
when $\Delta m \Delta t >> 2\pi$, where $\Delta t$ is the uncertainty of the
time measurement, the oscillation part is wiped out and the two species seem 
to be $45^0$ mixed.
\vskip.1in
 
Finally, let us calculate the loop diagrams in Fig. 1 in
order to estimate the size of the effects. 
The boson and fermion in the loops
of Fig. 1 can be $(\Delta^-/\Delta^0, l^+/\nu)$,  or
$(\phi^+/\phi^0, l^-/\nu)$,
  and the  first  combination has a potential to contribute
a significant effect.  
It has been realized  that the Higgs-fermion couplings may 
be large, 
since the discovery  of the top quark. As an example, the 
first combination will 
be considered here. For   right-handed neutrinos, in the basis 
where the right- 
handed neutrino masses are already diagonalized at the tree level,
\beq
\alpha_{12}=\frac{m_{R1}m_{R2}}{V_R^2}
	           \sum_{i}^{}
												{\it\bf M}_{i}^R {\it\bf I_i}
\eeq
where 
            ${\it\bf M}_{i}^R=(U^{l}_{1i}U^{l*}_{2i})_{_R}$,
	      ${\it\bf I_i}_R=\frac{1}{16\pi^2}\frac{m_{l_i}^2}
	      {m_{_{R2}}}
(-1  - i\pi\frac{m_{l_i}^2}{m_{R2}^2})$ 
and $\mu_{_R}$ is the mass of $\Delta_R$. The calculation
is scale ($P^2)$  sensitive because  the  outside propagator
is $\frac{\not P + m_{_{R2}}}{P^2-m_{R2}^2}$. 
The uncertainty in the  momentum flow $(P^2)$ will disappear in special 
physical situations. $P^2$ 
is chosen to be $P^2=\frac{1}{2}(m_{R1}^2+m_{R2}^2)$ in (27) 
and the assumption of
$P^2>>\mu_R^2+m_l^2$ is made.

\vskip.1in
The diagrams in Figure 1 for left-handed neutrinos do not
 enjoy GIM suppression, and  therefore are
 divergent.
  A more 
careful calculation is needed in order to obtain loop corrections.  
A guess is that because of the smallness of the left-handed 
neutrino masses, which are produced by the see-saw mechanism, the 
corrections
can be large to satisfy the condition for {\bf Case 2}, perhaps for 
one pair of the left-handed neutrinos.  

\vskip.1in
Some of these discussions may apply to models with both
Dirac and Majorana masses for charged particles[9]. In these models,
opposite charged leptons  coexist.
\vskip.1in 
 The FCNC and Majorana interactions among neutrinos
 may play a role in the neutrino
scattering in the early universe when the temperature is very high.
These interactions do not respect lepton number  conservation,
e.g. one may have 
$$\nu_{_R} +\nu_{_R}\rightarrow \Delta_R^o\rightarrow \bar\nu_{_R} +
\bar\nu_{_R}.$$
This process is possible because $\Delta_R$, which is a component
of the right-handed triplet, develops VEV.  $\Delta_R$ is part 
of the 126-plet
in the $SO(10)$ models (in some models, 126 is a composite field).
These interactions provide a  vehicle for lepton and anti-lepton 
numbers to reach an equilibrium at extremely high temperatures, 
even if there is a large lepton  number excess 
at the beginning. On the other hand, the existence of CP 
violation in the 
neutrino sector plus lepton number
nonconservation processes 
may contribute to a baryon number excess immediately after the
 decoupling of some heavy particles[6].  
\vskip.1in
In conclusion, massive neutrino models are likely to have 
flavor changing
neutral currents in the Higgs mediated neutrino interactions.  
In the
models that use the see-saw mechanism to explain the smallness of the
neutrino masses, FCNC are inevitable.  
\vskip.1in
The author acknowledges R. Arnowitt for conversations and 
discussions
at an early stage of this work and Z.Z. Xing for 
communications and comments. He thanks X.M. Zhang and H.Q. 
Zheng and D. Lichtenberg for encouragement.
This work began during a summer visit to Texas A\&M University. 
The National Science Foundation
partially supported this work by a NSF HRD grant.
\clearpage
\begin{figure}
\begin{center}
\epsfig{file=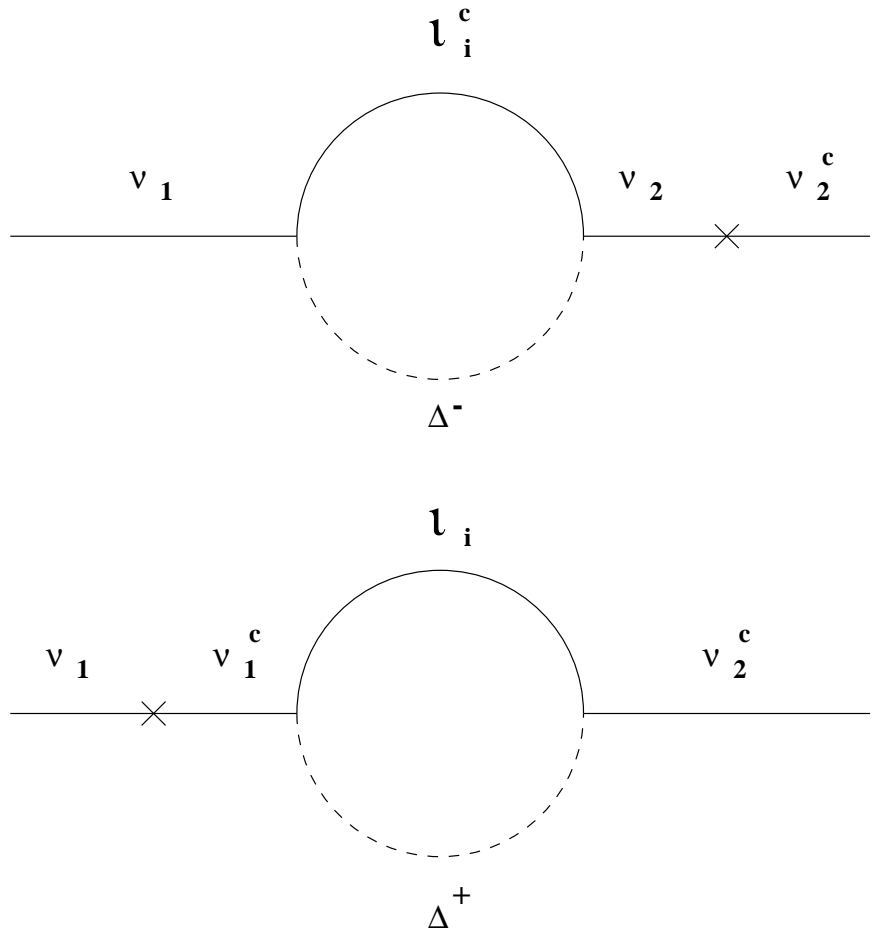}
\caption{ A one loop contribution to the mass matrix}
\end{center}
\end{figure} 
 
\clearpage
{\bf References}
\begin{enumerate}
\item 
B. Pontecovo, Zh. Eksp. Teor. Fiz. 33, (1957) 549; F. Nezrick 
and F. Reines,
Phys. Rev. 142 (1966) 852.
 For a recent review see, e.g. 
 M. Narayan, M.V.N.
Murthy, G. Rajaskaran, S. Uma Sankar, Phys. Rev. D 53 (1996) 
2809;
 see also, G.L. Gogli, E. Lisi, D. Montanino, and G. Scioscia, 
 hep-ph/9607251;
LSND Collaboration, Las Alamos Bulletin Board, hep-ph/9605001.

\item M. Gell-Mann, P. Ramond, and R. Slansky, in 
{\it Supergravity}, ed.
D. Freedman and P. van Niuenhuizen, (North Holland, 1979); 
T. Yanagida, KEK Proceedings
(1979); R. N. Mohapatra and G. Senjanovic, Phys. Rev. Lett. 
44 (1980) 912.

\item H. Georgi, Proc. AIP, Ed. C.E. Carlson, Meeting at 
William \& Mary 
College, 1974; H. Fritzsch and P. Minkowsky, Ann. Phys. 
(NY) 93 (1975) 193;
 M. Gell-Mann, P. Ramond, and R. Slansky, Rev. Mod. Phys. 
 50 (1978) 721;
S. Rajpoot, and P. Sithikong, Phys. Rev. D22 (1980) 2244;
   F. Wilczek and A. Zee, Phys. Rev. D25 (1982) 553; J.A. 
   Harvey, D. B. 
Reiss, and P. Remond, Nucl. Phys. B199 (1982) 223; R.N. 
Mahapatra, {\it
Unification and Supersymmetry}, Springer-Verlag, 1986;   
D.D. Wu and Y.L. Wu, PVAMU-HEP-12-95 and hep-ph/9603418, 
Mod. Phys. Lett. A11 (1996) 2703.
\item D.D. Wu, Nucl. Phys. B199 (1982) 533.
\item D.D. Wu, Phys. Lett. B90B (1980)451.
 D.D. Wu. Phys. Rev. D33 (1986) 860R; D.D. Wu and Y.L. Wu, 
Chinese Phys. Lett. 4 (1987) 441.
\item See, e.g. T. Yanagida, and M. Yoshimura, Phys. Rev. 
D23 (1981) 2048.
\item See e.g. M. Flanz, E. A. Paschos, U. Sarkar. and J. 
Weiss, Los Alamos
Bulletin Board,
hep-ph/9607310 and its citations, Phys. Lett. B, forthcoming.
\item
S. Glashow, J. Iliopoulos, and L. Maiani,
Phys. Rev. D2 (1960)1285; A. de Rujula, and S. Glashow, Phys. 
Rev. Lett.
45 (1980) 942.
\item D.D. Wu and T.Z. Li, Nucl. Phys. B245 (1984) 532.
\end{enumerate}
\end{document}